\documentclass[aps,pra,twocolumn,showpacs,notitlepage,floatfix,superscriptaddress]{revtex4-1}
\usepackage{amsfonts}
\usepackage{mathtools}
\usepackage{graphicx}
\usepackage{epsfig}
\usepackage{dcolumn}
\usepackage{bm}
\usepackage{amsmath}

\usepackage[normalem]{ulem}
\usepackage[latin1]{inputenc}
\usepackage{ulem}
\usepackage{epstopdf}
\usepackage{subfigure}
\usepackage{color}
\usepackage[dvipsnames]{xcolor}
\usepackage{amsthm}
\usepackage{newlfont}
\usepackage{graphicx}
\usepackage{amssymb}
\usepackage{epstopdf}
\usepackage{appendix}
\usepackage[breaklinks=true]{hyperref}
\usepackage{breakcites}
\usepackage{textcomp}
\usepackage{appendix}
\usepackage{multirow}	
\usepackage{color}
\usepackage{amssymb}
\usepackage{epsfig}
\usepackage{bm}
\usepackage[american]{babel}
\usepackage{braket}
\hypersetup{colorlinks=true,linkcolor=blue,citecolor=blue,filecolor=blue,urlcolor=blue,pdfstartview=FitH}

\begin{document}
\title{Statistical mechanics of one-dimensional quantum droplets}
\author{T. Mithun}
 \affiliation{Department of Mathematics and Statistics, University of Massachusetts, Amherst MA 01003-4515, USA}
 
 \author{S. I. Mistakidis}
\affiliation{Center for Optical Quantum Technologies, Department of Physics, University of Hamburg, 
Luruper Chaussee 149, 22761 Hamburg Germany}

 \author{P. Schmelcher}
\affiliation{Center for Optical Quantum Technologies, Department of Physics, University of Hamburg, 
Luruper Chaussee 149, 22761 Hamburg Germany} \affiliation{The Hamburg Centre for Ultrafast Imaging,
University of Hamburg, Luruper Chaussee 149, 22761 Hamburg,
Germany}

\author{P. G. Kevrekidis}
 \affiliation{Department of Mathematics and Statistics, University of Massachusetts, Amherst MA 01003-4515, USA}

\begin{abstract}
We study the statistical mechanics and the dynamical relaxation process of modulationally unstable one-dimensional quantum droplets described by a modified Gross-Pitaevskii equation. To determine the classical partition function thereof, we leverage the semi-analytical transfer integral operator (TIO) technique. The latter predicts a distribution of the observed
wave function amplitudes and yields two-point correlation functions providing insights into the emergent dynamics involving quantum droplets. 
We compare the ensuing TIO results with the probability distributions obtained at large times of the modulationally unstable dynamics as well as with the equilibrium properties of a suitably constructed Langevin dynamics.  
We find that the instability leads to the spontaneous formation of quantum droplets featuring multiple collisions and by which are found to coalesce at large evolution times. Our results from the distinct methodologies are in good agreement aside from the case of low temperatures in the special limit where the droplet widens. In this limit, the distribution acquires a pronounced bimodal character,  exhibiting a deviation between the TIO solution and the Langevin dynamics.
\end{abstract}
\maketitle

\section{Introduction} \label{intro}

The celebrated Gross-Pitaveskii model has  proved particularly successful for studying and  describing a variety of macroscopic many-body phenomena in
zero-temperature Bose-Einstein condensates (BECs)~\cite{pethick2008bose,stringari,siambook}.  
On the other hand, more recently, a new type of matter wave has emerged \cite{petrov2015quantum,Petrov_2016}, 
namely the so-called quantum droplets. The theoretical basis is a two-component (binary) BECs,  with
intra-component self-repulsion, yet also inter-component attraction lying in the vicinity of the self-repulsion. Here, the
famous Lee-Huang-Yang quantum correction~\cite{lee1957eigenvalues}
comes into play to account for the averaged effect of quantum fluctuations beyond the mean-field description. It competes with the mean-field
effects~\cite{pethick2008bose,stringari} and can prevent the possibility of BEC collapse present in the mean-field realm. 
Notice that this stabilization mechanism depends crucially on the system's dimensionality; namely beyond mean-field fluctuations are attractive in one-dimension (1D) while being repulsive in higher spatial dimensions. Importantly, these predictions led to a sequence of experimental realizations of such quantum 
droplets~\cite{cabrera2018quantum,cheiney2018bright,Semeghini2018self,Ferioli_2019,fort}, firstly observed in dipolar gases~\cite{Chomaz2016dipolar,Ferrier2016dipolar}.  
Accordingly, these states constitute nowdays an emerging topic that is under intensive research in the context of BECs.  

Along this vein of experimental developments,
it is remarkable that collisions of such droplet patterns
have been experimentally observed recently (leading to mergers
for slow collisions and quasi-elastic passage for fast
ones)~\cite{Ferioli_2019}. Additionally and while the above examples
focused on $^{39}$K droplets, heteronuclear binary BECs of $^{87}$Rb and $^{41}$K showcased very stable quantum droplets on time scales of
the order of a second~\cite{fort}. 
In parallel to this steady stream of experimental demonstrations,
theoretical studies have spearheaded a number of parallel directions.
These include but are not limited to the exploration of {self-evaporation dynamics of droplets \cite{Ferioli2020dynamical,app11020866}, quantum
droplets with intrinsic vorticity~\cite{yongyao}, the impact of
discreteness in the form of semidiscrete droplets
with or without topological charge~\cite{semidiscrete},
or the investigation of 3D stable generalizations of such
states~\cite{kartashov}.
Many of these developments have been recently summarized in Ref.~\cite{luo2020new}. 
Interestingly, the 1D dynamical features of droplet states are far less explored and are currently mainly restricted to inelastic collisional aspects of these configurations especially for high momenta and flat-top droplets~\cite{Astrakharchik_2018} or their spontaneous generation due to the modulation instability (MI)~\cite{mithun2019inter}. {Additionally, the phase diagram of quantum droplets trapped in 1D optical lattices was studied very recently~\cite{Morera2020QD}.}  

The above efforts have mainly focused on the dynamical aspects of the
droplets, and principally so at the zero temperature setting. 
However, naturally,
studying the finite temperature dynamics of BEC systems is a topic of
broad
theoretical and experimental appeal~\cite{griffin2009bose}. 
{It is indeed challenging to completely eliminate thermal effects in current experiments, a fact that further justifies the usage of models that can operate at finite temperatures~\cite{DeRosi2021thermal}. }
This is particularly intriguing in 1D where the role of quantum fluctuations, being inherently related to droplet formation, is more prominent~\cite{parisi2019liquid,parisi2020quantum,mistakidis2021formation} and three-body losses are suppressed~\cite{astrakharchik2006correlation} compared to
higher dimensions while the experimental probe of these configurations is still elusive. 
One of the commonly used methods to include the temperature induced fluctuations is the so-called truncated Wigner method \cite{blakie2008dynamics,PhysRevA.101.043604}. Recently, a
considerable
amount of attention has been drawn towards the so-called
positive P-method too \cite{deuar2006first,deuar2006firsta}. In
general, the final task of these methods is solving a stochastic
equation that incorporates the thermal fluctuations to the
corresponding Hamiltonian dynamical system. 
   In statistical mechanics, a principal task consists of the
   evaluation of the partition function corresponding to the energy
   functional of the system at hand. A central tool to this effect in
   1D settings consists of the so-called transfer integral operator
   (TIO) method~\cite{scalapino1972statistical}. Indeed, this method
   renders the above problem equivalent to solving the single particle
   Schr\"odinger equation.  Moreover, it has been shown that a
   stochastic method represented by a suitable Langevin equation is
   comparable with
   the solution of TIO \cite{khare1997exact, Nunnenkamp_2007,PhysRevLett.84.1070}. Further, recently it has been shown that the steady state of molecular dynamics 
   and the classical field method  also reproduces the TIO results \cite{Nunnenkamp_2007,aarts2000thermalization}. 

 To the best of our knowledge, such techniques motivated by 
 statistical
 mechanics, while being successful at the level of single-component
 Gross-Pitaevskii model~\cite{Nunnenkamp_2007}, have yet to be explored in the quantum
 droplet realm of effectively multi-component systems. Indeed, this is a central focus of
 the present work.
More concretely, we study the statistical mechanics of a quantum
droplet in a 1D ring geometry. 
One of the peculiarities of a quantum droplet is its incompressibility leading to a maximal critical density for sufficiently large particle numbers. 

This relevant system, as mentioned above, can be represented
with a modified binary Gross-Pitaveskii (MGP) equation, where the form of a
Lee-Huang-Yang term is determined by the dimension of the system 
\cite{Petrov_2016,Astrakharchik_2018,PhysRevA.101.051601,lavoine20201d}. For the case of particle-balanced components with 
equal masses and equal intra-component interaction
strengths, which will be of primary interest herein,
the binary MGP equation reduces to a single-component one. We determine the classical partition function corresponding
to our model by mapping the functional integration of the partition
function to a single-particle Schr\"odinger equation via
the TIO technique. 
{Note that the classical partition function corresponds to the high-temperature limit of the quantum partition function describing an interacting Bose gas \cite{pippard1964elements}.}
We then numerically verify the predictions of TIO via a suitably crafted Langevin dynamics \cite{khare1997exact,
  Nunnenkamp_2007}. Our results indicate that the equilibrium properties of the Langevin dynamics are well in line with the TIO solution at {intermediate} and high temperatures for $\mu \rightarrow \mu_0$,
the limit where the droplets tend to disappear; the agreement is less adequate at {low} temperatures. 
We also compare the TIO results with the long-time dynamical
  evolution of
  the original droplet system in the regime where the latter falls
  into MI~\cite{nguyen2017formation} and
  subsequently  relaxes. The recent studies on the statistical
  properties of MI motivate
  this analysis~\cite{PhysRevLett.123.234102,PhysRevLett.123.093902}.
  Indeed, in Ref.~\cite{mithun2019inter}, it has been shown that the
  MI resulting from the small perturbation of a plane-wave state leads
  to the formation of quantum droplet structures that undergo
  inelastic collisions.  
  Interestingly,  we find that, as a result of the inelastic
  collisions these generated droplet structures coalesce and their
  equilibrium properties are well matched by the TIO analysis.  
  It is worthwhile to 
  mention here that within our study we neglect any possible temperature dependence of the Lee-Huang-Yang term, a prospect that remains still an open question.

 The work is organized as follows.  We introduce the MGP model in
 Sec. \ref{sec1} and then determine the classical partition function
 by using the TIO  method in Sec.~\ref{sec2}. Section \ref{sec3} is
 devoted to
 developing and exploring the relevant Langevin dynamics. We report the results of the MI dynamics within the MGP framework in Sec.~\ref{sec5} and provide an outlook in Sec.~\ref{sec6}. {Appendix~\ref{TIO_deri} elaborates on the extraction of the single-particle Schr\"odinger picture through the transfer integral method. Appendix~\ref{2comp_lang} explicates the derivation of the Langevin equations in interaction imbalanced two-component mixtures.  
 {Finally in Appendix~\ref{temp_background} it is shown that the inclusion of thermal effects in the mixture, emulated by a dissipative term in the MGP, leads to a more pronounced coalescence process of the droplets associated with the suppression of the background fluctuations.} }
 
\section{The modified Gross-Pitaevskii framework} \label{sec1}

{The starting point of our analysis will be the  dimensionless MGP equation describing 1D quantum droplets emerging in symmetric binary mixtures~\cite{Petrov_2016,mithun2019inter}. In particular,
\begin{equation}
i\frac{\partial \psi }{\partial t}=-\frac{1}{2}\frac{\partial ^{2}\psi }{%
\partial z^{2}}+|\psi |^{2}\psi -|\psi |\psi-\mu\psi,  \label{eq:1GP}
\end{equation}%
with the normalization condition
\begin{equation}
\int_{-\infty }^{+\infty }n~dz=N,~~~\text{and}~~~n=|\psi (z)|^{2}.  \label{eq2}
\end{equation} 
Here, $N$ is the number of atoms and $\mu$ represents the chemical potential. The effective single-component Eq.~(\ref{eq:1GP}) is obtained as a reduction of the respective coupled set of two-component modified Gross-Pitaveskii equations [see also Appendix~\ref{2comp_lang}] under symmetry considerations. The latter refer to a mixture with same particle number per species ($N_1=N_1\equiv N/2$),  equal repulsive intra-component interaction strengths ($g_1=g_2\equiv g>0$) and equal masses ($m_1=m_2\equiv m$), see for details~\cite{Petrov_2016,Astrakharchik_2018}.  
Moreover according to Eq.~(\ref{eq:1GP}), the units of length, time and wave function are expressed in terms of the healing length $\xi$, $\hbar/(m \xi^2)$, and $(2 \sqrt{g})^{3/2}/\sqrt{\pi \xi} (2 |\delta g|)^{3/4}$, respectively, with $\xi=\frac{\pi \hbar^2}{m}\frac{\sqrt{2|\delta g|}}{g^{3/2}}$.
Also, $ \delta g =g_{12}+g$ and $g =\sqrt{g_1 g_2}$,  where 
$g > 0$ represent the repulsive intra-component interaction strengths {of the symmetric mixture}, while $g_{12}<0$ denotes the inter-component attractive interaction. Additionally,  for the existence of a quantum droplet $\delta g \ll g$ should hold. In a corresponding experiment, $\delta g$ can be tuned using the Feshbach resonance technique~\cite{cabrera2018quantum,cheiney2018bright}.

For $\delta g/g>0$, Eq.~\eqref{eq:1GP} gives rise to an exact localized flat-top (FT) solution~\cite{Astrakharchik_2018}. 
The latter represents a quantum droplet which originates from the balance between the effective cubic self-repulsion and quadratic attraction characterized by the central density $n_{0}$ and the chemical potential $\mu _{0}$, where
 \begin{equation}
 n_{0}=\frac{4}{9},~~\mu _{0}=-\frac{2}{9},   \label{1n0}
 \end{equation}
 respectively. 
 In what follows we shall focus on the $\delta g /g>0$ regime. 
 However, for reasons of completeness we remark that solutions to Eq. ~(\ref{eq:1GP}) exist also for $\delta g/g<0$, see Ref.~\cite{mithun2019inter} for a more elaborated discussion.

 \section{Statistical Mechanics} \label{sec2} 
 
  To study the statistical mechanics of the model under consideration, as described by 
   Eq.~\eqref{eq:1GP}, we establish the corresponding classical partition function via the TIO method,
  as indicated in the Introduction in line with Refs.~\cite{Nunnenkamp_2007,scalapino1972statistical,khare1997exact}.  For completeness, we will present selected results for the two-component case in Appendix B. In order to proceed, we consider the droplet on a periodic domain (i.e., effectively
  a ring) of length $L$. Then,  the form of the classical partition function
  reads
 \begin{equation}
Z=\int D(\psi,\psi^{\ast})e^{-\beta F[\psi,\psi^{\ast}]},
\label{parti1}
\end{equation}%
 where 
 \begin{equation}
 \begin{split}
F&=\int dz(H-\mu N)
\label{free_ener}
\end{split}
\end{equation}%
 is the free energy, $\beta$ is the inverse temperature, $H$ represents the Hamiltonian corresponding to Eq.~\eqref{eq:1GP} and $N$ is the total number of particles. 
 In accordance with the TIO methodology, the functional integration can be reduced to an eigenvalue problem \cite{scalapino1972statistical,PhysRevB.22.5522}. For the Eq.~\eqref{eq:1GP}, we arrive at the corresponding eigenvalue equation (see Appendix \ref{TIO_deri} for a detailed derivation)
 \begin{equation}
\Bigg[-\frac{1}{2\beta^2}\frac{\delta ^{2} }{%
\delta \psi^{2}}+V_d(\psi)\Bigg]\phi_n(\psi) = E_n \phi_n(\psi).
\label{tio1}
\end{equation}%
Evidently, Eq.~(\ref{tio1}) corresponds to a 1D single-particle Schr\"odinger equation.
In this expression, $\phi_n$ represent the corresponding single-particle eigenfunctions, being functionals of the $\psi$ field, while $E_n$ denote the eigenvalues. Importantly, the effective TIO potential appearing in Eq.~(\ref{tio1}), possesses the form
\begin{equation}
 V_d(\psi) = \frac{1}{2}|\psi|^{4}-\frac{2}{3}|\psi|^{3} - \mu |\psi|^2.
\label{tio2}
\end{equation} 
Notably, this represents the common mean-field potential of the Gross-Pitaevskii theory
\cite{Petrov_2016}.  
The shape of these effective potentials in the different parameter regions of the system is crucial in order to understand where bound state solutions, and thus droplet-like configurations, are prone to appear. 
For this reason, we next show $V_d(\psi=u)$ in Fig.~\ref{fig:apot} since the one-component case will be our focus in the following.
According to Eq.~(\ref{tio2}) the minimum of $V_d(\psi=u)$ is given by 
$u=(3 b + \sqrt{9 b^2 + 32 a \mu})/(8a)$ for $\mu > 0$, while $u=(3 b \pm \sqrt{9 b^2 + 32 a \mu})/(8a)$ for $\mu \le 0$ where $a=\frac{1}{2}$ and $b=\frac{2}{3} $.
As a result for $\mu<0$ the effective potential features a double-well structure, see Fig.~\ref{fig:apot}. This feature is related to the presence of a bimodal probability distribution as it will be argued later on [Eq.~(\ref{eq:pda})].     

\begin{figure}[!htbp]
\centering
\includegraphics[width=0.89\linewidth]{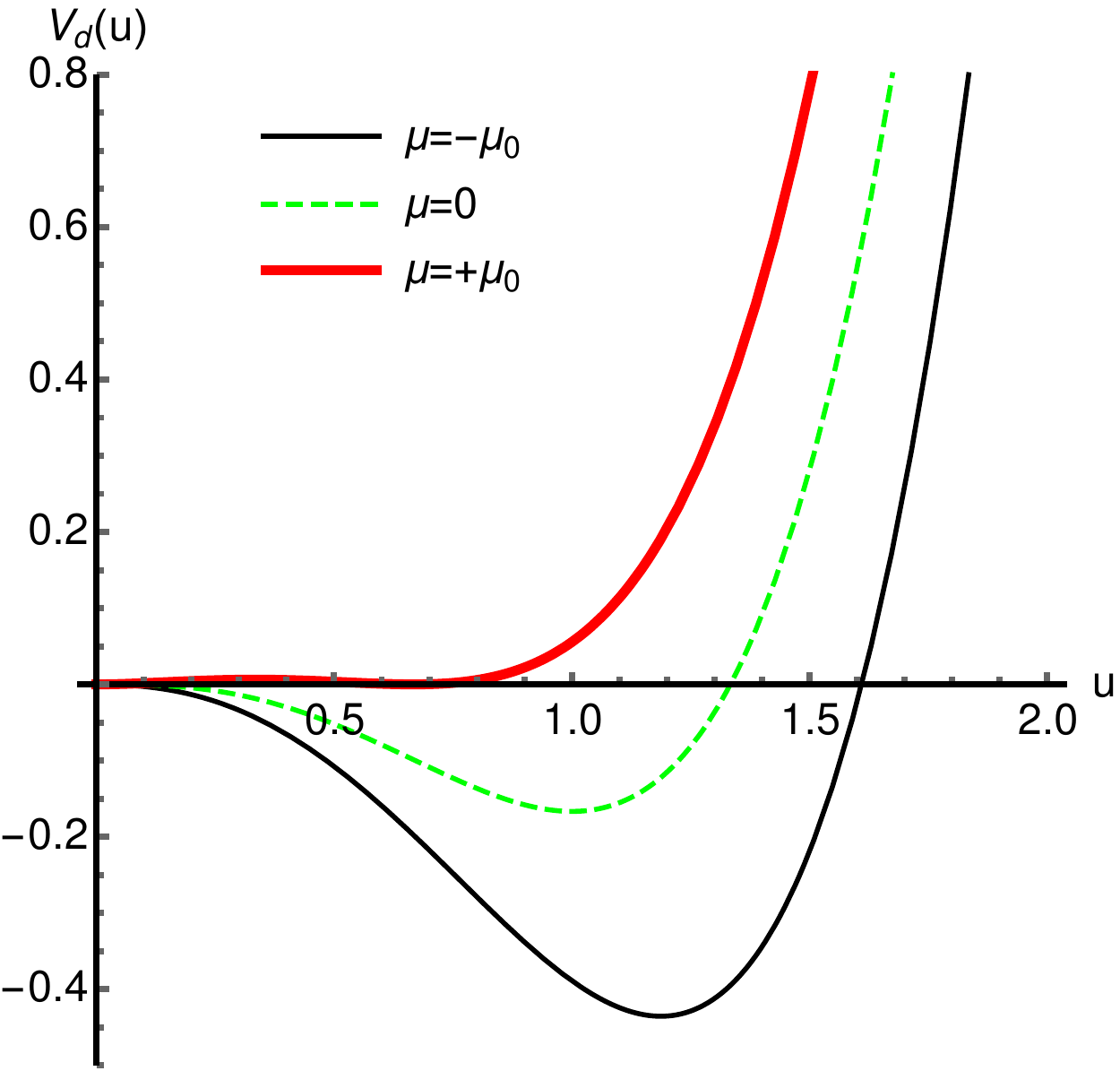}
\caption{Effective anharmonic potential $V_d(u)$, described by Eq.~(\ref{tio2}).  
Shown are different values of the chemical potential $\mu$ (see legend) with $\mu_0$ given by Eq.~(\ref{1n0}). }
\label{fig:apot}
\end{figure}

We solve the Eq.~\eqref{tio1} numerically to determine the corresponding eigenvalues and eigenvectors utilizing exact diagonalization.  Notice that this is the only aspect that renders the TIO approach semi-analytical. 
The main contribution to the partition function in Eq.~(\ref{parti1}) will be from the lowest TIO eigenvalue. 
The thermodynamic properties of an equilibrium system can be well understood by resorting to the underlying probability distribution of amplitudes (PDA), $P(|\psi|=u)$. 
The latter can be expressed in terms of the eigenfunction corresponding to the lowest eigenvalue $E_0$ of Eq.~(\ref{tio1})~\cite{Nunnenkamp_2007}.  In particular $P(|\psi|=u)$ acquires the form 
 \begin{equation}
 P(|\psi|=u)=2 u|\phi_0(u)|^2.
\label{eq:pda}
\end{equation}
Moreover, the steady state properties of a thermodynamic system can be further characterized by employing the two-point spatial correlation function $C(z)$~\cite{naraschewski1999spatial,katsimiga2017dark}. 
Here, $z$ denotes the relative distance between two distinct spatial locations. 
This correlation function can be written with respect to the eigenvalues and eigenfunctions of Eq.~(\ref{tio1}) as 
  \begin{equation}
  \begin{split}
 C(z)&=\langle\psi(.)\psi(.+z) \rangle \\&= \sum_n\lvert\int du \phi_n^{\ast}(u)u\phi_0(u)\rvert^2e^{-\beta|z|(E_n-E_0)}.
\label{eq:cor}
\end{split}
\end{equation}
It provides a measure of the coherence among the particles~\cite{naraschewski1999spatial,mistakidis2018correlation}.  
Concretely, it is bounded from above and below taking as a maximal and  minimal value the considered particle number and zero respectively. 
If $C(z)$ is maximal for every $z$ then the system is termed fully coherent and it is characterized by quasi long-range order. 
Otherwise losses of coherence occur. 
In the thermodynamic equilibrium, the long-range order is expected to vanish and accordingly $C(z)$ tends to zero for increasing $z$~\cite{nandkishore2015many}. 
We finally remark that, for the settings considered herein, the contributions of the (higher) excited states of the effective potentials (described by Eqs.~(\ref{eq:pda}) and (\ref{eq:cor})) are neglected, as being exponentially smaller than the dominant lowest-lying states. 
\section{Langevin dynamics}\label{sec3} 
It has been demonstrated that, for any positive temperature, the exact results obtained from the TIO in a BEC system can be compared with that of a Langevin equation with a Gaussian additive white noise term~\cite{Nunnenkamp_2007}. 
This type of stochastic dynamics contributes towards relaxing the
system configuration to the free-energy minimum, while at the same time
accounting for the thermal fluctuations arising around this minimum \cite{parisi1981perturbation}.  
Indeed, the point of the Langevin dynamics is to allow the fields to evolve
in a way such that they eventually relax to the appropriate 
equilibrium distribution and then to sample
the suitable observables. I.e., this
dynamics relaxes the field configuration
via fictitious time-evolution to the relevant minimum
of the free energy surface, while the white
noise randomly drives the field.  
In the following, we consider the Langevin equations corresponding to  Eq.~(\ref{eq:1GP}),  in line also with the earlier work of~\cite{Nunnenkamp_2007}
for regular one-component BECs
\begin{equation}
\frac{\partial \psi }{\partial t}=-\Big[-\frac{1}{2}\frac{\partial ^{2}}{%
\partial z^{2}}+|\psi |^{2} -|\psi| -\mu \Big]\psi+\xi_1(z,t).  \label{eq:lg1}
\end{equation} 
The term $\xi_1$ represents the white noise having a Gaussian distribution with the correlation 
\begin{equation}
\langle\xi_1^{\ast}(z,t)\xi_1(z',t')\rangle = \frac{2}{\beta}\delta(z-z')\delta(t-t').  \label{eqlg2}
\end{equation}%
The Langevin equations are indeed found to relax to the equilibrium state at large evolution times. This is caused by the fact that the time average of the spatio-temporal correlation relaxes to its equilibrium spatial correlation or in other words the time average of the noise approaches its equilibrium distribution \cite{Hohenberg_1977}. We solve Eq.~\eqref{eq:lg1} numerically by using the xmds package \cite{dennis2013xmds2}. 
\begin{figure}[!htbp]
\centering
\includegraphics[width=0.98\linewidth]{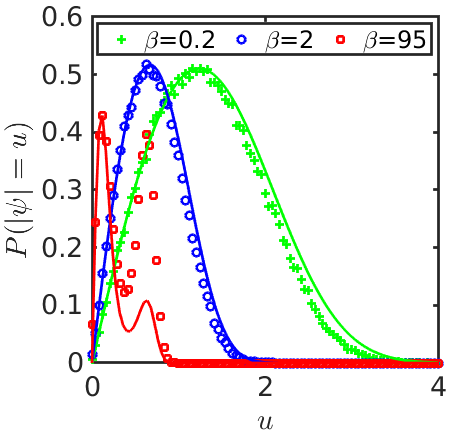}
\caption{Probability distribution function obtained from TIO (solid lines) and within the Langevin dynamics for different values of the inverse temperature $\beta$. Namely, $\beta=0.2$ (green crosses), $\beta=2$ (blue circles) and $\beta=95$ (red squares). {The chemical potential used is $\mu=\mu_0+0.00001$.}}
\label{fig:tio}
\end{figure}
\begin{figure}[!htbp]
\centering
\includegraphics[width=0.48\textwidth]{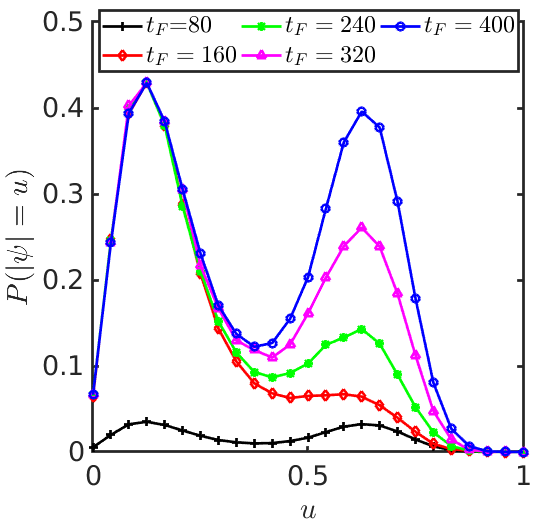}
\caption{Probability distribution function $P(|\psi|=u)$ at different total evolution times $t_F$ (see legend) for $\beta=95$ obtained within the Langevin dynamics. 
A saturation of $P(|\psi|=u)$ is observed for $t_F>200$. 
{The chemical potential used is $\mu=\mu_0+0.00001$. }}
\label{fig:tioa}
\end{figure}

Note that for the numerical simulations of Eq.~(\ref{eq:lg1}), to be presented below, a sample of 1000 trajectories is found to be sufficient for the convergence of the relevant probability distributions, with the domain size being $L=30$. 
We first fix the chemical potential $\mu=\mu_0+0.00001$, where the exact solution of Eq.~\eqref{eq:1GP} is a FT shaped {droplet}. {Figure~\ref{fig:tio} depicts the probability distribution $P(|\psi|=u)$ results of the Langevin dynamics for 
(long) total evolution time $t_F \ge 200$.} The green (crosses), blue (circles) and red (squares) symbols represent the numerically obtained  steady states for $\beta=0.2$,  $\beta=2$ and  $\beta=95$, respectively; the solid lines represent the corresponding TIO solutions. 
Inspecting $P(|\psi|=u)$ [Fig.~\ref{fig:tio}] we can readily deduce that as the temperature decreases (from $\beta=0.2$ to $\beta=95$), it changes from a single peak distribution (unimodal) to a bimodal one in terms of the TIO prediction.  
The comparison between the TIO results and the ones within the Langevin approach highlights that the steady state of the Langevin dynamics is in  very good agreement with the TIO solutions both at higher and {intermediate} temperatures. On the other hand, at the {low} temperature regime, the steady state of the Langevin dynamics deviates from the TIO solution. The latter case is the more numerically delicate one 
due to the nature of the relevant effective TIO potential, a feature to 
which we attribute the observed discrepancy. 
In order to confirm that this discrepancy is not due to the considered total evolution time $t_F$ that allows to reach the steady state, we determine the $P(|\psi|=u)$ for different $t_F$ at $\beta=95$, see in particular Fig.~\ref{fig:tioa}.  
Interestingly, we observe that the probability distribution for evolution times in the range of $t_F\in (160-240$) develops a bimodal structure, being proximal to the expected picture from the TIO analysis.  
However, upon considering larger timescales the relative difference in the amplitude between the humps appearing in $P(|\psi|=u)$ tends to be suppressed. 

We will further address this discrepancy when considering the dynamics of the full model, Eq.~(\ref{eq:1GP}), (rather than the modified Langevin one) in the following section. 
To gain additional insight on the dependence of the steady state distribution at low temperatures, we now vary the chemical potential $\mu$  by fixing the inverse temperature parameter $\beta=95$. The obtained results for the $P(|\psi|=u)$ are provided in Fig.~\ref{fig:beta}. It shows that the TIO solution develops a bimodal distribution {\it only when} $\mu \rightarrow \mu_0$. As $\mu$ deviates
from this FT {droplet} limit, the TIO solution leads to a unimodal distribution and the results of the Langevin dynamics match well with these solutions. 
This once again suggests that presumably the Langevin dynamics is not
able to capture the delicate FT configuration of the {droplet} at the intermediate
temperature limit, as we will
argue further below. Nevertheless, in all other settings, the TIO
semi-analytical prediction is adequately captured by the modified
Langevin dynamics.
\begin{figure}[!htbp]
\centering
\includegraphics[width=0.98\linewidth]{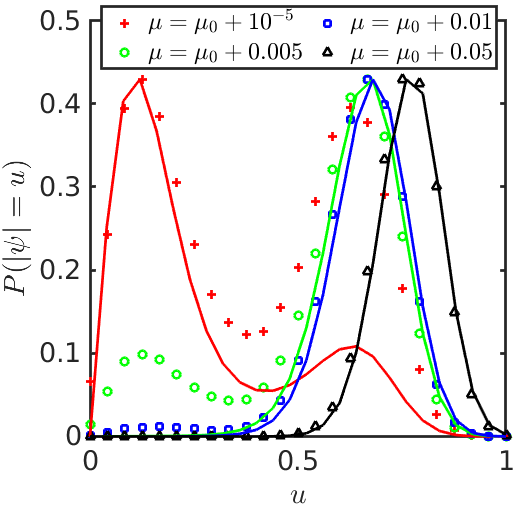}
\caption{Probability distribution $P(|\psi|=u)$ obtained from the Langevin dynamics for $\mu=\mu_0+0.00001$ (red crosses), $\mu=\mu_0+0.005$ (green circles), $\mu=\mu_0+0.01$ (blue squares),  and $\mu=\mu_0+0.05$ (black triangles). The solid lines represent the corresponding TIO solutions. The inverse temperature is $\beta=95$.}
\label{fig:beta}
\end{figure}

To gain further insight into the emergent steady state of the {droplets}, we additionally determine the two-point correlation function $C(|z|)$ [Eq.~(\ref{eq:cor})]. 
The latter is provided in Fig.~\ref{fig:tioo} for sufficiently large evolution times, i.e., $t_F\geq 200$. This observable allows to assess the underlying coherence losses which are induced here by the temperature. It evinces that the predictions of the TIO are in accordance with the ones of the Langevin dynamics for both $\beta=0.2$ and $\beta=2$ inverse temperatures [see lower and upper panels of Fig.~\ref{fig:tioo}].  Particularly, quasi long-range order is suppressed and therefore coherence is lost since $C(|z|)\to 0$ for increasing $z$. 
This fact also supports the appearance of a steady state~\cite{nandkishore2015many} for the droplet.  
On the other hand, at the large inverse temperature ($\beta =95$), in line with the mismatch in the probability distributions, the behavior of the correlation function deviates from the respective transfer integral prediction in the 
$\mu \rightarrow \mu_0$ limit as seen in the lower panel of Fig.~\ref{fig:tioo}. 
Concretely, the TIO method shows a vanishing coherence for larger values of $z$, whilst the Langevin dynamics predicts that the coherence of the system is maintained.  
\begin{figure}[!htbp]
\centering
\includegraphics[width=0.99\linewidth]{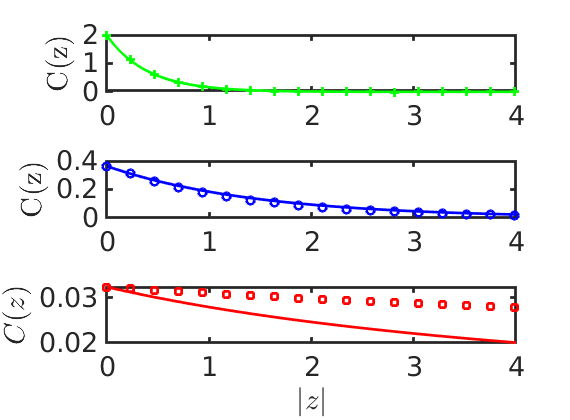}
\caption{Correlation function $C(z)$ obtained from the TIO method (solid lines) and from the Langevin dynamics (dots) for different temperatures. Namely, $\beta=0.2$ (top panel), $\beta=2$ (middle panel) and $\beta=95$ (bottom panel) at $\mu=\mu_0+0.00001$. Here the considered total evolution time is $t_F \ge 200$.}
\label{fig:tioo}
\end{figure}

\section{MI dynamics in the modified Gross-Pitaevskii  approach} \label{sec5}

Recently, it has been shown that the dynamics of a classical field method 
(CFM) adequately traces the features of the TIO solution
for the equilibrium statistical mechanics of the single-component BEC case~\cite{Nunnenkamp_2007}. 
Indeed, a key point of that work is the comparison of three different
methods: the Langevin approach detailed above with a molecular dynamics (MD)
one~\cite{aarts2000thermalization} (based on Hamiltonian dynamics of Klein-Gordon type) and the CFM.
Considering the MD methodology as less directly related to the system
at hand, we focus here on the CFM as a complement of the Langevin approach
presented above.
The core idea of the CFM is to obtain a steady state where the dynamics is governed by the Gross-Pitaevskii equation and then compare this steady state solution with the TIO solution. In this method, the initial state is a non-equlibirium one where only a few modes in momentum space are excited. Moreover, it is well known that in such systems the MI phenomenon naturally creates non-equilibrium conditions when subjected to a small perturbation (seeding the relevant instability)~\cite{nguyen2017formation,everitt2017observation}.  
The Gross-Pitaevskii equation with an effective attractive interaction is known to be modulationally unstable. The MI dynamics of the model as described by Eq.~(\ref{eq:1GP})
has been recently studied in detail~\cite{mithun2019inter} and argued to lead to droplet nucleation. 
Motivated by this finding we shall subsequently focus on the dynamical response of the symmetric mixture generated by its direct time-evolution within the MGP and being associated with the unstable dynamics stemming
from the MI. In particular, we aim to unravel the spontaneous creation of droplets due to MI and their fate in the long-time dynamics. 
Indeed, we allow the system to evolve into
its asymptotic (equilibrium) state from an initially dynamically 
unstable initial condition and ``observe'' the resulting 
effective temperature by comparison with the TIO solution results. 

\begin{figure}[!htbp]
\centering
\includegraphics[width=1\linewidth]{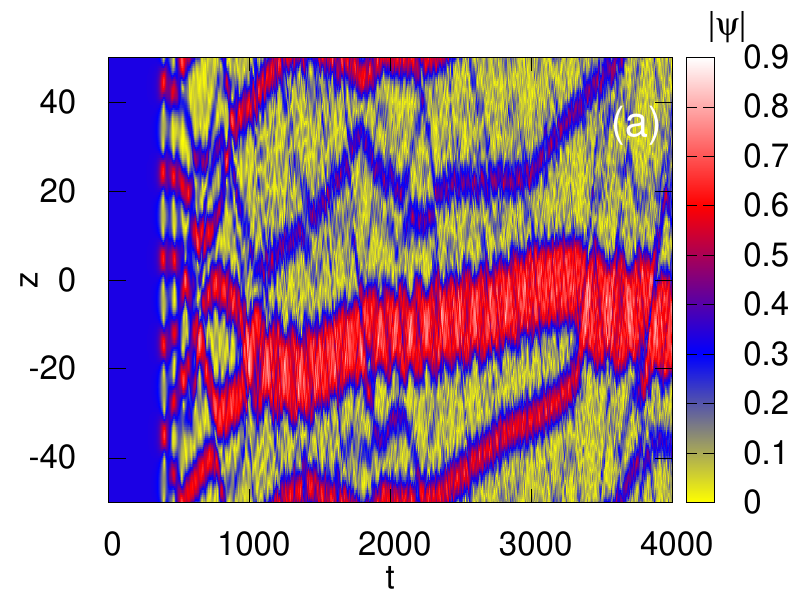}
\includegraphics[width=0.93\linewidth]{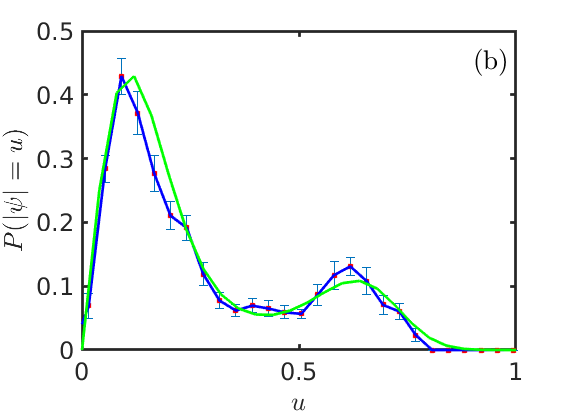}
\includegraphics[width=1\linewidth]{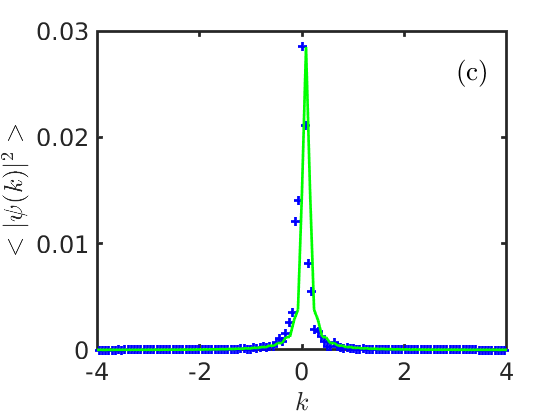}
\caption{MI dynamics for the system parameter values
$n\approx0.11$ and $\mu=\mu_0+0.00001$. (a) Spatiotemporal evolution of $|\psi(z,t)|$. 
(b) Probability distribution function $P(|\psi|=u)$ and (c) correlation function {in momentum space} $\langle |\psi(k)|^2\rangle$ at a final evolution time of
$t_F=4000$.  The results shown in (b) and (c) are averaged ones over a sample of twelve different initial conditions and the error bars shown in (b) represent the respective standard deviation.  
The green solid line represents the TIO solution at $\beta=95$, while, the blue dotted points are obtained through the direct simulation of the MGP as described by Eq.~(\ref{eq:1GP}). }
\label{homodyn_compare}
\end{figure}

As an initial condition we consider a homogeneous particle density distribution subjected to a weak amplitude perturbation of the form $\delta \psi = A_p  e^{i \theta_r}$. 
The strength $A_p$ of the perturbation is of the order of $10^{-8}$ and $\theta_r$ represents the uniform noise distribution taking values in the interval $[0,\pi]$~\cite{mithun2019inter}. 
We fix $\mu = \mu_0+0.00001$ that corresponds to  $n\approx 0.11$. The results are presented in Fig.~\ref{homodyn_compare}. The MI initially leads to the formation of small droplets as discussed in \cite{mithun2019inter}. In the course of the evolution these small droplets undergo inelastic collisions 
(analogously to what was observed in~\cite{Ferioli_2019}) and
feature a coarsening stage. This results in the  formation of ``large'' {droplets} as shown in Fig.~\ref{homodyn_compare}(a). Also, we noted that the amplitude of these waveforms is close to the central density, $n_0$ of the FT shaped droplet. The corresponding PDA is depicted in  Fig.~\ref{homodyn_compare}(b) (blue dotted points). Interestingly, the PDA exhibits a bimodal distribution, where the peak at large amplitude ($u\approx 0.66$) represents the large amplitude droplets, while peaks at small amplitude ($u\approx 0.33$) are related to small-amplitude phononic excitations~\cite{PhysRevA.101.051601} that are widespread within the spatial extension of the cloud. 

To compare this equilibrium state with the TIO solution, we considered
different values of $\beta$ and obtained that, e.g., for $\beta=95$ the probability distribution functions are in close correspondence with each other as shown in Fig.~\ref{homodyn_compare}(b) with the green solid line. Interestingly, the accordance between the PDA and TIO solution predictions holds also for the corresponding correlation function in momentum space~\cite{mistakidis2015resonant} i.e. {$\braket{|\psi(k)|^2}=(1/2\pi) \int dz C(z) e^{-ikz}$. The latter is routinely accessible in current ultracold atom experiments via time-of-flight imaging ~\cite{bloch2008many}. 
This observable is provided in Fig.~\ref{homodyn_compare}(c) at $t_F=4000$ and $\beta=95$. As can be seen, it features a central peak structure around $k=0$ signaling the steady state of the {droplet} while its shape is in very good agreement between the two approaches.} To further corroborate that the same argumentation holds also for other values of the chemical potential, we exemplarily showcase in Fig.~\ref{MI2} the cases of $\mu=\mu_0+0.01$ and $\mu=\mu_0+0.1$. 
{It becomes apparent that even for these values of the chemical potential, the final state resulting from the MI dynamics is a large excitation on top of a fluctuating background bearing small amplitudes. 
Hence, the corresponding PDA is a bimodal distribution. 
However, the weight of the small amplitude excitations in the corresponding PDA distribution decreases with increasing $\mu$, in line with the respective TIO prediction shown in Fig.~\ref{fig:beta}. } 

In our considerations so far, we have limited the
study of an effective temperature and chemical potential
associated with the free energy and atom number of our
droplet forming system. In Appendix~\ref{temp_background}, we also briefly consider
a variant of the MGP dynamics involving dissipative perturbations,
as a first glimpse towards the role of finite temperature effects
in the spontaneous generation of droplets.
In particular, we explicate that the  above-discussed coalescence process of droplets becomes more prominent in the presence of thermal effects of the background as captured by the direct inclusion of a damping term in the modified Gross-Pitaevskii equation. 

\begin{figure}[!htbp]
\centering
\includegraphics[width=1\linewidth]{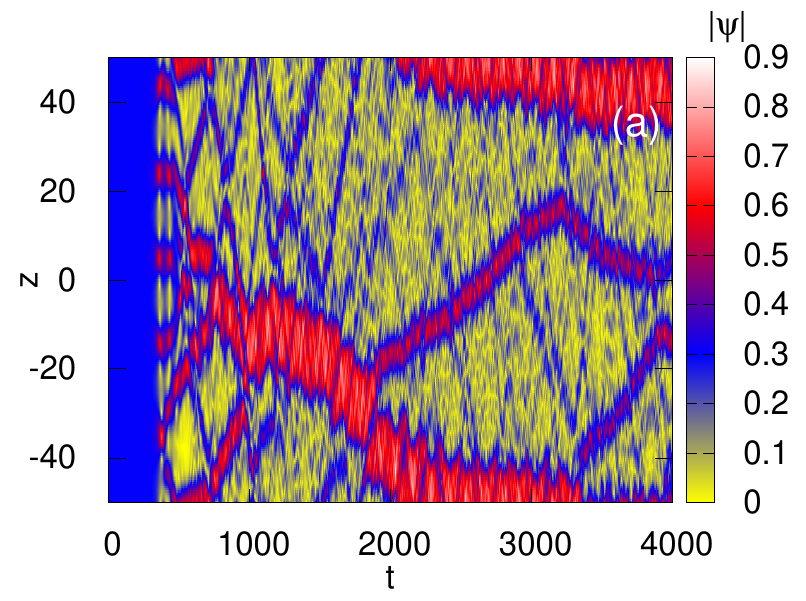}
\includegraphics[width=1\linewidth]{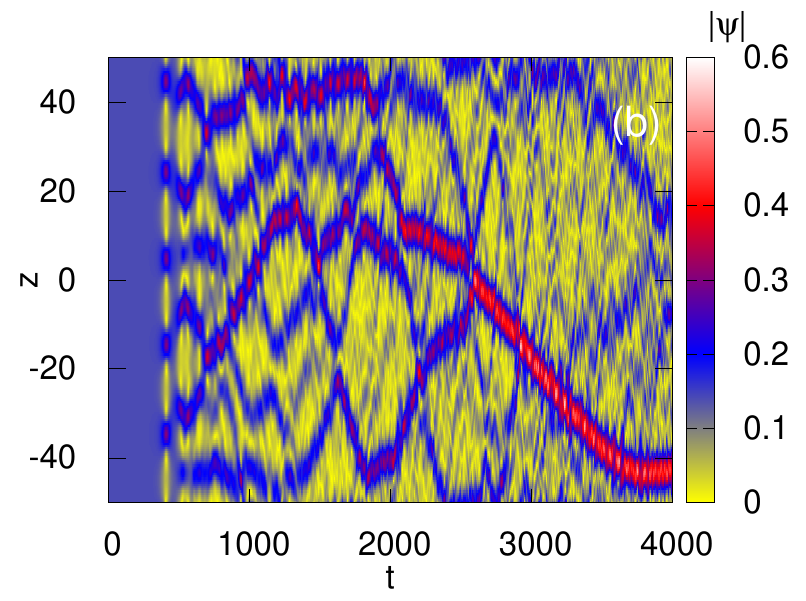}
\caption{Time-evolution of $|\psi(z,t)|$ explicating the MI dynamics, the subsequent generation of {droplets} and their coalescence while (a) $\mu=\mu_0+0.01$, and (b) $\mu=\mu_0+0.1$.}
\label{MI2}
\end{figure}

\section{Conclusions and Future Challenges}\label{sec6}
 
  In the present work, we have sought to address the equilibrium statistical mechanics of one-dimensional quantum droplets.  This system is represented by a modified two-component Gross-Pitaevskii equation and under the symmetry considerations (equal wave functions and equal intra-component interaction strengths {and masses}) it reduces to a single-component equation.  To determine the classical partition function, we first reduce the functional integration of the partition function to a single-particle Schr\"odinger problem by using the TIO technique.  The probability distribution of the amplitudes obtained from TIO shows  unimodality in the limit $\mu \rightarrow \mu_0$ at intermediate and large temperatures.  
  However, at small temperatures the relevant probability exhibits a bimodal shape. On the other hand,  for the values of $\mu$ far from $\mu_0$, the  probability 
 distribution is always found to be unimodal. 
 
  We compared the results of the TIO analysis with the equilibrium properties  of a suitably crafted Langevin dynamics.  The equilibrium properties of the Langevin dynamics reproduce well the corresponding ones of the TIO solutions except in the limit $\mu \rightarrow \mu_0$ for low temperatures. 
  In this latter scenario, a bimodal distribution is featured showing in turn a partial discrepancy between the equilibrium properties of the Langevin dynamics and the TIO solutions.  
  Additionally,  we compared the probability distribution of the modulational instability dynamics at large times with the TIO findings, obtaining good  agreement in the cases under consideration for a suitable choice of the temperature.  
  Importantly, we have shown that the modulational instability inherent to the system gives rise to spontaneous quantum droplet nucleation. 
  These in turn undergo a coarsening stage experiencing collision events and leading to their coalescence.  
 
  Very recently, significant attention has been drawn to the thermodynamics of quantum droplets {\cite{DeRosi2021thermal}}. Beyond the realm of the present
  study, several topics remain open for future studies. 
  An extension of thermodynamic considerations in higher dimensional
  settings would be a topic of particular interest for a variety of
  reasons. On the one hand, semi-analytical tools such as the TIO
  do not straightforwardly generalize to higher dimensions, hence 
  different theoretical approaches would need to be brought to bear.
  Moreover, even numerically the nonlinearities of the system being
  logarithmic in 2D and featuring a different form than the one
  considered herein ($\propto |\psi|^3 \psi$) in 3D could yield
  different outcomes as regards the probability distributions and
  the correlation functions. On the other hand, the genuinely 
  multi-component case where the two components are not forced
  to be equal and its systematic consideration and direct comparison e.g. with a variational approach~\cite{mistakidis2018correlation} is also of interest for further study. 
  
 \section*{Acknowledgments}
We thank S. Flach and K. Burnett for fruitful comments and discussions. 
S. I. M. gratefully acknowledges financial support in the framework of the Lenz-Ising Award of the 
University of Hamburg. The present paper is based on work that was supported by the US National Science Foundation under DMS-1809074 and PHY-2110030 (PGK).

\appendix
 \section{Derivation of the transfer integral problem}\label{TIO_deri}
 Here, we briefly discuss the derivation of the single-particle Schr{\"o}dinger Eq.~\eqref{tio1}, describing the equilibrium state of the symmetric mixture, from the underlying classical partition function of Eq.~\eqref{parti1}~\cite{scalapino1972statistical}.  
 For a ring of length { $L=M\Delta z$, where $M$ represents the number of grid points and $\Delta z$ denotes the (quite fine) spatial discretization used in the numerical implementation of
 the problem}, we can 
 discretize the free energy functional introduced in Eq.~\eqref{free_ener} as
  \begin{equation}
  \begin{split}
F&=\Delta z \sum_{i=1}^M f(\psi_i,\psi_{i+1})=
\Delta z \sum_{i=1}^M\Big[\frac{1}{2}|\frac{\psi_{i+1}-\psi_i}{\Delta z}|^2\\&+\frac{1}{2}n_i^{2}-\frac{2}{3}n_i^{3/2} - \mu n_i\Big],
\end{split}
\label{free_enera}
\end{equation}%
 and we obtain the partition function \cite{scalapino1972statistical}
  \begin{equation}
Z= \prod_{i=1}^M \int_{-\infty}^{\infty} d\tilde{\psi}_1^{'} d\tilde{\psi}_i e^{-\beta \Delta z f(\psi_i,\psi_{i+1})}\delta(\tilde{\psi}_1-\tilde{\psi}_1^{'}),
\label{part_ap1}
\end{equation}%
 where $d\tilde{\psi}_i = \sqrt{\frac{\beta }{2 \pi \Delta z}} d\psi_i$ and $M$ is the number of segments within the ring of width $\Delta z$. 
 We now expand the $\delta$ function in terms of a normalized set of (complete) eigenfunctions and obtain
   \begin{equation}
Z= \sum_n \prod_{i=1}^M \int_{-\infty}^{\infty}  d\tilde{\psi}_1^{'} d\tilde{\psi}_i \phi_n(\tilde{\psi}_1^{'}) e^{-\beta \Delta z f(\psi_i,\psi_{i+1})}\phi_n(\tilde{\psi}_1).
\label{part_ap2}
\end{equation}%
 We can calculate this partition function from the eigenvalue problem 
    \begin{equation}
 \int_{-\infty}^{\infty} d\tilde{\psi}_i e^{-\beta \Delta z f(\psi_i,\psi_{i+1})}\phi_n(\tilde{\psi}_i)=e^{-\beta \Delta z E_n}\phi_n(\tilde{\psi}_{i+1}),
\label{part_ap3}
\end{equation}%
 where $\phi_n$ and $E_n$ are the eigenfunctions and eigenvalues of the   transfer matrix  equation. We then reduce this equation to the single-particle Schr{\"o}dinger problem as follows. Performing the Taylor series expansion of  the eigenfunction $\phi_n(\tilde{\psi}_i)$ around $(\tilde{\psi}_{i+1})$ we arrive at  
 \begin{widetext}
    \begin{equation}
    \begin{split}
 \int_{-\infty}^{\infty} d\tilde{\psi}_i e^{-\beta \Delta z f(\psi_i,\psi_{i+1})}\phi_n(\tilde{\psi}_i)=e^{-\beta \Delta z \Big[\frac{1}{2}\frac{\delta g}{g}n^{2}-\frac{1}{2} \frac{2^{5/2}}{3\pi }n^{3/2} - \mu n \Big]} &\times \Big(1+\frac{\Delta z}{2 \beta }\frac{\partial^2}{\partial \psi_{i+1}^2}\Big)\phi_n(\tilde{\psi}_{i+1})\\&
 =e^{-\beta \Delta z H } \phi_n(\tilde{\psi}_{i+1})\\& \equiv e^{-\beta \Delta z E_n } \phi_n(\tilde{\psi}_{i+1})
     \end{split}
\label{part_ap4}
\end{equation}%
\end{widetext}

where $\Big(1+\frac{\Delta z}{2 \beta }\frac{\partial^2}{\partial \psi_{i+1}^2}\Big)\approx  e^{\frac{\Delta z}{2 \beta }\frac{\partial^2}{\partial \psi_{i+1}^2}}$.

The effective Hamiltonian reads

\begin{equation}
    \begin{split}
 H=-\frac{1}{2 \beta^2 }\frac{\partial^2}{\partial \psi^2}+ \frac{1}{2}n^{2}-\frac{2}{3}n^{3/2} - \mu n.
     \end{split}
\label{part_ap5}
\end{equation}%
Here, the second term corresponds to the mean-field contribution and the third one is the leading-order beyond mean-field correction accounting for quantum fluctuations. 

\begin{figure}[!htbp]
\centering
\includegraphics[width=0.98\linewidth]{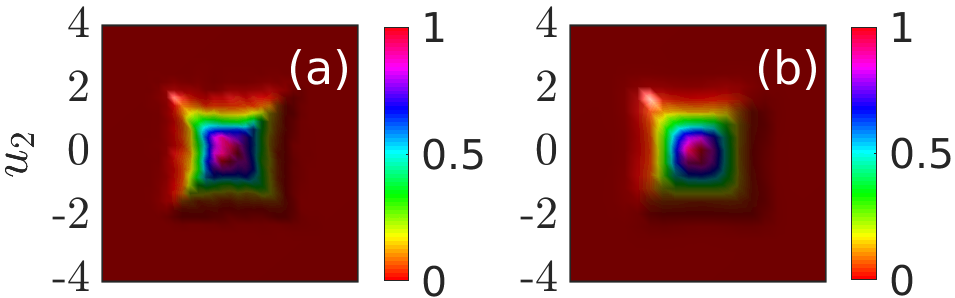}
\includegraphics[width=0.98\linewidth]{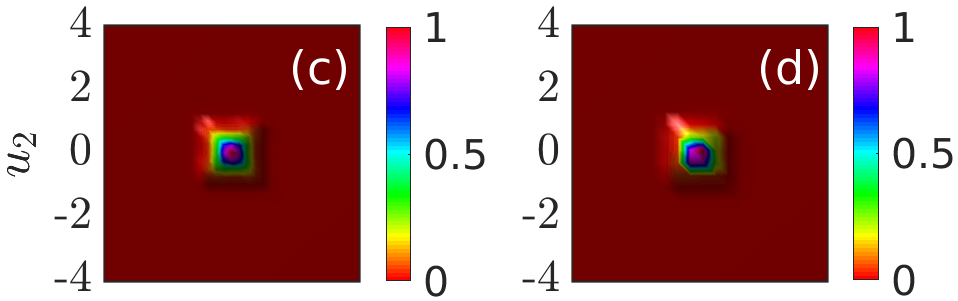}
\includegraphics[width=0.98\linewidth]{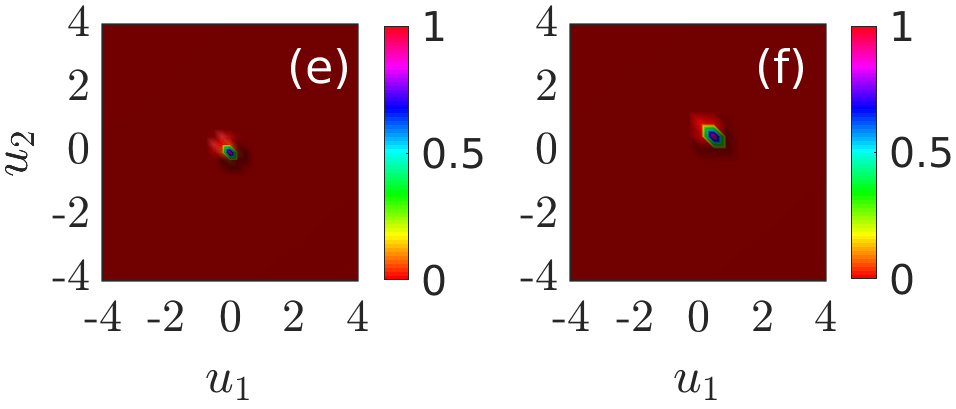}
\caption{Two-dimensional probability distribution function $P[{\textrm Re}(\psi_1)=u_1,{\textrm Re}(\psi_2)=u_2]$ obtained from the Langevin dynamics (left panels) and the TIO approach (right panels) for $\beta=0.2$ (top panels), $\beta=2$ (middle panels) and $\beta=95$ (bottom panels). 
Other parameters used are $\mu=\mu_0+0.00001$, $\delta g/g=0.05$, and $\mathcal{P}=1$. 
The total evolution time is $t_F = 240$.}
\label{fig:lg2a}
\end{figure}
\begin{figure}[!htbp]
\centering
\includegraphics[width=0.98\linewidth]{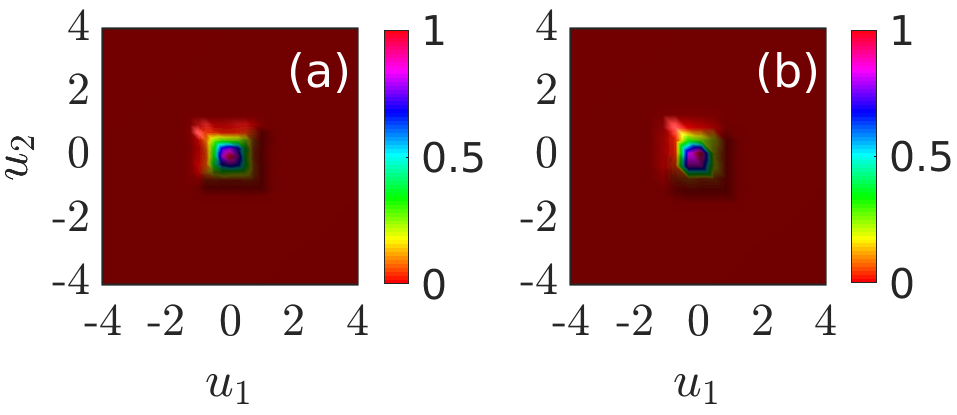}
\caption{Two-dimensional probability distribution function $P[Re(\psi_1)=u_1,Re(\psi_2)=u_2]$ as predicted  from the Langevin dynamics (left panel) and the TIO method (right panel) for $\beta=2$, $\mathcal{P}=1.25$ as well as $\mu=\mu_0+0.00001$. 
The considered total evolution time corresponds to $t_F = 240$.}
\label{fig:lg2b}
\end{figure}

\section{Two-component Langevin equations}\label{2comp_lang}
{Having analyzed in the main text the case of symmetric mixtures, here we  briefly discuss the generalization of our method to interaction imbalanced two-component settings.
For this reason we resort to the  dimensionless MGP equation for a 1D binary quantum droplet~\cite{Petrov_2016,mithun2019inter},
 \begin{equation}
\begin{split}
i\frac{\partial \psi _{1}}{\partial t}& =-\frac{1}{2}\frac{\partial ^{2}\psi
_{1}}{\partial z^{2}}+\mathcal{Q} (\mathcal{P}+\mathcal{G}\mathcal{P}^{-1})|\psi _{1}|^{2}\psi _{1}-\mathcal{Q}(1-\mathcal{G})|\psi
_{2}|^{2}\psi _{1}\\&-\mathcal{P}\sqrt{\mathcal{P}|\psi _{1}|^{2}+\mathcal{P}^{-1}|\psi _{2}|^{2}%
}\psi _{1}-\mu_1 \psi_1, \\
i\frac{\partial \psi _{2}}{\partial t}& =-\frac{1}{2}\frac{\partial ^{2}\psi
_{2}}{\partial z^{2}}+\mathcal{Q}(\mathcal{P}^{-1}+\mathcal{G}\mathcal{P})|\psi _{2}|^{2}\psi _{2}-\mathcal{Q}(1-\mathcal{G})|\psi
_{1}|^{2}\psi _{2}\\&-\frac{1}{\mathcal{P}}\sqrt{\mathcal{P}^{-1}|\psi _{2}|^{2}
+\mathcal{P}|\psi _{1}|^{2}}\psi _{2}-\mu_2 \psi_2,
\end{split}
\label{eq:2GP}
\end{equation}%
where the involved parameters stand for
\begin{equation}
\begin{split}
\mathcal{P}\equiv \sqrt{\frac{g_{1}}{g_{2}}},&~\mathcal{Q}=\frac{g(1+\mathcal{P})^2}{ 2\mathcal{P} \delta g},~\mathcal{G}= \frac{2\mathcal{P}^2 \delta g}{g(1+\mathcal{P}^2)^2} \\&~\text{and}~ \delta g =g_{12}+g.  \label{PG}
\end{split}
\end{equation}
In the above expressions, $g_1 > 0$ and $g_2 > 0$ correspond to the repulsive intra-component interaction strengths, while $g_{12}<0$ is the inter-component attractive interaction. Also, $\mu_1$ and $\mu_2$ are the chemical potentials of the individual subsystems. The parameter $\mathcal{P}$ quantifies the intra-component interaction imbalance and $\mathcal{G}$ measures the deviation from the balance point of the mean-field repulsion and attraction where $\delta g=0$. 
In the coupled system of Eqs.~(\ref{eq:2GP}) the units of length, time and the wave function are expressed in terms of $\xi$, $\hbar/(m \xi^2)$, and $\frac{(\sqrt{g_1}+\sqrt{g_2})^{3/2}}{\sqrt{\pi \xi}(2|\delta g|)^{3/4}}$, with the healing length being $\xi=\frac{\pi \hbar^2}{m}\frac{\sqrt{2|\delta g|}}{g(\sqrt{g_1}+\sqrt{g_2})}$.}

{For the Eq.~\eqref{eq:2GP}, we arrive at the corresponding eigenvalue equation
 \begin{equation}
 \begin{split}
\Bigg[-\frac{1}{2\beta^2}\Big(\frac{\delta ^{2} }{%
\delta \psi_1^{2}}+\frac{\delta ^{2} }{
\delta \psi_2^{2}}\Big)&+V_{2d}(\psi_1,\psi_2)\Bigg]\phi_n(\psi_1,\psi_2) \\&=E_n \phi_n(\psi_1,\psi_2),
\label{tio11a}
\end{split}
\end{equation}
As it can be seen,  Eq.~(\ref{tio11a}) is a two-dimensional single-particle Schr\"odinger equation.
Also, $\phi_n$ refer to the single-particle eigenfunctions, being functionals of the $\psi$ field, and $E_n$ are the respective eigenvalues. 
The effective TIO potential of Eq.~(\ref{tio11a}) reads
\begin{equation}
\begin{split}
V_{2d}(\psi_1,\psi_2)&=\mathcal{Q}\frac{\mathcal{P}+\mathcal{G} \mathcal{P}^{-1}}{2}|\psi_{1}|^{4}+\mathcal{Q}\frac{\mathcal{P}^{-1}+\mathcal{G}\mathcal{P}}{2}|\psi_{2}|^{4}\\&+\mathcal{Q}(\mathcal{G}-1) |\psi_{1}|^{2} |\psi_{2}|^{2}-\frac{2}{3}\left( \mathcal{P}|\psi_{1}|^{2}+\frac{|\psi_{2}|^{2}}{\mathcal{P}}\right) ^{3/2}\\&-\mu (|\psi_{1}|^{2}+|\psi_{2}|^{2}).
\end{split}
\label{zeropressurecond}
\end{equation}%
Below, we consider the Langevin equations corresponding to  Eqs.~(\ref{eq:2GP}), namely
\begin{equation}
\begin{split}
\frac{\partial \psi _{1}}{\partial t}& =-\Big[-\frac{1}{2}\frac{\partial ^{2}\psi
_{1}}{\partial z^{2}}+\mathcal{Q}(\mathcal{P}+\mathcal{G}\mathcal{P}^{-1})|\psi _{1}|^{2}\psi _{1}\\&-\mathcal{Q}(1-\mathcal{G})|\psi
_{2}|^{2}\psi _{1}-\mathcal{P}\sqrt{\mathcal{P}|\psi _{1}|^{2}+\mathcal{P}^{-1}|\psi _{2}|^{2}%
}\psi _{1}\\&-\mu_1\psi_1\Big]+\xi_1(z,t), \\
\frac{\partial \psi _{2}}{\partial t}& =-\Big[-\frac{1}{2}\frac{\partial ^{2}\psi
_{2}}{\partial z^{2}}+\mathcal{Q}(\mathcal{P}^{-1}+\mathcal{G}\mathcal{P})|\psi _{2}|^{2}\psi _{2}\\&-\mathcal{Q}(1-\mathcal{G})|\psi
_{1}|^{2}\psi _{2}-\frac{1}{ \mathcal{P}}\sqrt{\mathcal{P}^{-1}|\psi _{2}|^{2}
+\mathcal{P}|\psi _{1}|^{2}}\psi _{2}\\&-\mu_2\psi_2\Big]+\xi_2(z,t).
\end{split}
\label{eq:lg2}
\end{equation}}
{For the numerical simulation of Eq.~(\ref{eq:lg2}) we utilize 1000 trajectories with the domain size $L=8\times8$. We first set $\mu_1 \approx \mu_2 \equiv \mu_0+0.00001$ and $\mathcal{P}=1$. Further, by considering two different noise distributions ($\xi_1$ and $\xi_2$) for the two individual components, we break the symmetry between them. Figure~\ref{fig:lg2a} depicts the results of the Langevin dynamics (left panels) and the TIO solutions (right panels). Comparing the corresponding 2D probability distribution functions $P[{\textrm Re}(\psi_1)=u_1,\textrm{Re}(\psi_2)=u_2]$ between the Langevin
 dynamics and the TIO findings for various inverse temperatures we deduce that they are in a good agreement besides the region of small temperatures. Particularly, as the temperature increases the distribution widens similarly to the single-component case.  However, at $\beta=95$, we observe a mismatch among the predicted distributions. Indeed, the center of the distribution within the TIO case is shifted from the origin $(0,0)$ in contrast to the Langevin scenario where it is strongly localized around the center.  
  Furthermore, in order to estimate the impact of an increasing population asymmetry we change $\mathcal{P}=1$ to $\mathcal{P}=1.25$. The corresponding PDAs for $\beta=2$ are provided in Fig.~\ref{fig:lg2b} both within the Langevin method and the TIO solution. 
 Evidently, the predictions of these approaches are quite proximal. }

\begin{figure}[!htbp]
\centering
\includegraphics[width=1\linewidth]{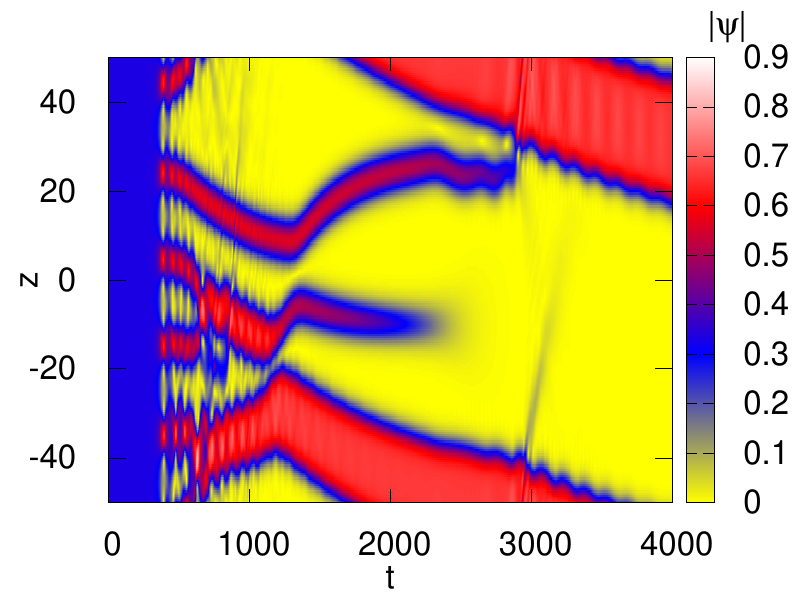}
\caption{Time-evolution of $|\psi(z,t)|$ in the presence of temperature showcasing the spontaneous creation of droplets due to the MI dynamics and their consequent coalescence. We use $\mu=\mu_0+0.00001$. The phenomenological factor accounting for the presence of temperature corresponds here to $\gamma=0.03$.}
\label{MI_dissi}
\end{figure}
\section{Phenomenological consideration of 
finite temperature effects towards droplet formation}\label{temp_background} 

Next, we briefly discuss the role of finite (condensate) temperature
effects in the spontaneous generation of the droplets, relying on certain approximations. 
It is important to recall that so far, we have been referring
to {\it effective} temperatures and chemical potentials, associated
with the (free) energy and atom number of the system in the 
grand-canonical ensemble formulation herein.
To emulate the role of thermal processes~\cite{Proukakis_2008},  we introduce a phenomenological damping factor $\gamma$ in the MGP equation by replacing $i \rightarrow (i-\gamma)$ in Eq.~(\ref{eq:1GP}).  
Recall that this is a widely used approximation in the context of the common Gross-Pitaevskii framework, i.e. in the absence of Lee-Huang-Yang contributions~\cite{Kasamatsu:2003:nonlinear, Choi:1998:phenomeno,Mithun:2016:disorder,katsimiga2021phase}. This factor accounts for the inclusion of thermal components in the condensate being usually present in cold atom experiments but it ignores any possible temperature dependence stemming from the Lee-Huang-Yang contribution. The temperature dependence of the latter term and its competition with the thermal condensate fraction are still open problems and certainly worth to be pursued in future studies. Since the total number of atoms is not conserved in the presence of $\gamma$, we treat the chemical potential as a time-dependent parameter and adjust it at each evolution time~\cite{Mithun_2018_Signature}. 

 Figure~\ref{MI_dissi} shows the time-evolution of $|\psi(z,t)|$ in the case of $\gamma=0.03$. 
Notice that in line with the discussion of Ref.~\cite{yan_prouk},
values of $\gamma$ within $0.00023--0.0023$ are relevant
for temperatures in the order of $10--100$nK. Nevertheless, 
we utilize a larger value here to
render apparent the fact that the nucleation of droplets takes place within the same timescale as compared to the $\gamma=0$ scenario [Fig.~\ref{homodyn_compare}(a)] and also their original number is the same.
However, it should also be noted that the presence of temperature enhances the formation of the "large" droplets by minimizing the background fluctuations, compare in particular Fig.~\ref{MI_dissi} and Fig.~\ref{homodyn_compare}(a). As mentioned above, a further systematic exploration of finite temperature condensates, including the role of such corrections to LHY terms is certainly a topic of particular interest for further study.
%
     \bibliographystyle{apsrev4}
\let\itshape\upshape
\normalem
\bibliography{reference1}

 \end{document}